\documentclass[apsrev,onecolumn,epsfig,tighten]{revtex4}
\usepackage[dvips]{graphicx}

\newcommand \beq{\begin{eqnarray}}
\newcommand \eeq{\end{eqnarray}}
\newcommand \bea{\begin{eqnarray}}
\newcommand \eea{\end{eqnarray}}
\newcommand \ga{\raisebox{-.5ex}{$\stackrel{>}{\sim}$}}
\newcommand \la{\raisebox{-.5ex}{$\stackrel{<}{\sim}$}}

\begin{document}

\title{Cold quantum gases: coherent quantum phenomena from Bose-Einstein
condensation to BCS pairing of fermions}

\author{Gordon Baym}
\affiliation{Department of Physics, University of Illinois at
Urbana-Champaign,
 1110 West Green Street, Urbana, IL 61801
 }
\date{\today}

\begin{abstract}

    Studies of trapped quantum gases of bosons and of fermions have opened up
a new range of many-body problems, having a strong overlap with nuclear and
neutron star physics.  Topics discussed here include:  the Bose yrast problem
-- how many-particle Bose systems carry extreme amounts of angular momentum;
the infrared divergent structure of the transition to Bose condensation in a
weakly interacting system; and the physics of extremely strongly interacting
Bose and Fermi systems, in the scale-free regime where the two body s-wave
scattering lengths are large compared with the interparticle spacing.  Such a
regime is realized experimentally through use of atomic Feshbach resonances.
Finally we discuss creation of BCS-paired states in trapped Fermi gases.

\end{abstract}

\maketitle

\section{INTRODUCTION}

    The past decade has seen remarkable developments in the study of the
physics of ultracold trapped quantum atom gases.  Below a certain critical
temperature, Bose systems become Bose-Einstein condensed superfluids, with a
macroscopic fraction of the particles occupying the lowest single particle
mode of the system.  In 1995 the first Bose-Einstein condensed gases of
bosonic alkali atoms were produced \cite{jila,mit,huletbec}.  In the past
several years considerable attention has focussed on cooling fermionic alkali
atoms to the point where they become degenerate \cite{brian,thomas-flow} and
indeed undergo BCS pairing \cite{jin,kinast,zwierlein,chin}.  These laboratory
phenomena in ultracold trapped quantum gases have a strong overlap with
nuclear and neutron star physics.  In this talk, I would like give an overview
of the field, and indicate why this area has become so challenging to nuclear
physicists.

    The ideas of Bose-Einstein condensation (BEC) have played an important
role over the years in understanding the properties of the vacuum and of
matter under extreme conditions.  Condensates of $\pi$ mesons \cite{picond}
and of K mesons \cite{kcond} have been studied as possible states of neutron
star matter.  Bosonic condensates, e.g., with non-vanishing expectation values
of quark operators, $\langle\bar u u\rangle$, $\langle\bar d d\rangle$,
$\langle\bar s s\rangle$, are fundamental features of the vacuum and the
structure of elementary particles such as the nucleon, and underly the
spontaneous breaking of the chiral symmetry of the strong interactions
\cite{chiral}.  One of the important aims of ultrarelativistic heavy ion
collisions at RHIC (and soon at LHC) is to produce chirally restored matter in
the form of quark-gluon plasmas; there one is asking the opposite of the
question asked in condensed matter physics, namely, what are the properties of
Bose-Einstein {\it de}-condensed matter \cite{rhic-inpc}.  Bose condensation
of Cooper pairs is familiar in nuclear physics, where BCS pairing explains
reduced moments of inertia of heavier nuclei.  Both the neutron and proton
components of neutron star matter, more generally, undergo BCS pairing to
become superfluid.  Such superfluidity most likely underlies the observed
glitches, or sudden rotational speedups, observed in some 30 pulsars to date
\cite{glitch,richard}.

    Experiments in atomic systems are done primarily on vapors of alkali metal
atoms.  The alkalis have one electron outside a close shell and an odd number
of protons; thus the statistics a given isotope obeys is governed by whether
the nucleus has an even number of neutrons, in which case the atom is a boson,
or odd, in which case it is a fermion.  Since odd-odd nuclei tend to be
unstable, most alkali atoms are bosons.  The principal actors in the study of
atomic Bose-Einstein condensates are $^{87}$Rb (half-life, $T_{1/2} = 4.75
\times 10^{10}$y) and $^{23}$Na; condensates of $^{7}$Li have also been
studied, but the effective low energy interactions between $^{7}$Li are
attractive, which limited the number of atoms that could be assembled in a
vapor.  Two fermionic alkalis are long enough lived to be readily used in
trapping experiments, $^{6}$Li, which is stable, and $^{40}$K, with $T_{1/2} =
1.3\times 10^9$y.  Typical clouds contain $\sim 10^6$ atoms, with density
$\sim 10^{14}$/cm$^3$, cooled via a combination of laser and then evaporative
cooling, to temperatures $\sim 10^{-8}$K, the coldest systems in the universe!

    Because the temperatures are so very low, with typical atomic velocites of
order mm/s, the two particle interatomic interactions are low energy s-wave,
described by a pseudopotential,
\beq
  v_{int}(\vec r_1-\vec r_2) = g\delta(\vec r_1-\vec r_2),
\eeq
where $g=4\pi\hbar^2a_s/m$, with $a_s$ the s-wave scattering length.  The
first generation of Bose-Einstein condensation experiments focussed on
phenomena described accurately within mean-field theory, via the
Gross-Pitaevskii equation \cite{leggett-rmp},
\begin{equation}
    i\hbar \frac{\partial}{\partial t} \Psi(\vec r\,,t) =
 -\frac{\hbar^{2}}{2m}\nabla^{2}\Psi(\vec r\,,t)
 + V(\vec r\,) \Psi(\vec r\,,t)
  +g|{\Psi(\vec r\,,t)}|^{2}\Psi(\vec r\,,t),
\end{equation}
for the condensate wave function, $\Psi(\vec r\,,t)=\langle \psi(\vec
r\,,t)\rangle$.  Here $\psi$ is the single particle annihilation operator, and
$V$ is the external trapping potential.  Such studies included the shape of
the condensate (primarily described by Thomas-Fermi theory) \cite{rb87}, the
elementary modes -- breathing, quadrupole oscillations \cite{stringmodes},
shorter wavelength sound propagation, and scissor modes, in which the atomic
cloud undergoes angular oscillations with respect an asymmetric trap, the
analogue of angular counter-oscillations of the neutron and proton
clouds in nuclei.

    Studies of two-body correlations,
$\langle\psi^\dagger\psi^\dagger\psi\psi\rangle$, through measurements of the
interaction energy, and three-body correlations
$\langle\psi^\dagger\psi^\dagger\psi^\dagger\psi\psi\psi\rangle$, through
measurements of rates at which three atoms ``recombine" into a molecule and
fast atom, provided direct evidence that the systems were Bose-condensed and
not simply condensed in space, e.g., in a normal weakly interacting gas,
\beq
  \langle\psi^\dagger\psi^\dagger\psi\psi\rangle =
  2\langle\psi^\dagger\psi\rangle^2,
\eeq
and
\beq
  \langle\psi^\dagger\psi^\dagger\psi^\dagger\psi\psi\psi\rangle
   =6\langle\psi^\dagger\psi\rangle^3,
\eeq
while in a condensate the factors of two and six are, as predicted and
observed, absent.  Noteworthy is the measurement of Ref.~\cite{bloch} of the
single particle correlation function $\langle \psi^\dagger(\vec r\,)\psi(\vec
r\,')\rangle$ in trapped $^{87}$Rb, which below the transition temperature
extends to a finite value at large $|\vec r - \vec r'|$, as predicted for a
Bose-condensed superfluid, but never actually observed in superfluid $^4$He.

    The first experiment to demonstrate directly the quantum coherence of Bose
condensates is that at MIT of Andrews et al.  \cite{andrews}.  This expeiment
cooled two independent atomic systems to below the transition temperature, and
then released them so that they expanded into each other, exhibiting quantum
interference, analogous to the interference of two classical electromagnetic
waves from independent sources.

    In the past several years, the field has undergone dramatic expansion, as
experiment has gone into new regimes, including strong coupling and trapped
fermions.  The enticement of trapping fermions is the possibility of producing
BCS paired superfluids of atoms.  Trapped bosons and fermions behave similarly
at high temperatures.  At low temperatures, condensed bosons fall to the
center of the trap, limited only by the uncertainty principle, while trapped
fermionics clouds below the Fermi degeneracy temperature, $T_f$, are supported
at larger radii by Fermi degeneracy pressure.  Current temperatures achieved
in Fermi systems are as low as a few percent of $T_f$ \cite{jin}, in systems
of some $10^6$ atoms.  Through rapid rotation, discussed below, one is able to
produce lattices of vortices, and study the Bose yrast problem, how
many-particle Bose systems carry extreme amounts of angular momentum.  By
means of optical lattices -- formed by counter-propagating lasers producing
standing electromagnetic waves -- one can control to an unprecedented degree
the environment in which the atoms sit.  For example, one can produce lattices
in one, two and three dimensions.  By increasing the lattice depth one can
change a system from a superfluid to a Mott insulator \cite{haensch2}.  By
means of Feshbach resonances in two particle scattering, discussed below, one
is able to control the strength and sign of the interaction between particles,
and thus dial the system into the strong coupling regime at will.  The latter
technique has enabled one to produce fermion systems in the BCS paired regime
and then follow their evolution, as the interaction is changed from attractive
to repulsive, into the regime where the atoms from a Bose condensate of
molecules.  Among other areas of present interest are production of coherent
mixtures of atoms and molecules, spinor gases, fragmented condensates, and
mixtures of bosons and fermions.

\section{RAPIDLY ROTATING BOSE CONDENSATES}

    Understanding how Bose-Einstein condensates carry extreme amounts of
angular momentum explores regimes of many-body physics not encountered in
other systems \cite{qfs}.  As one rotates a Bose condensate rapidly, it forms
a triangular array of singly quantized vortex lines
\cite{Abo,HaljanCornell,jilatk,jila3,coddington}.  The angular momentum of the
system is $\sim N_v N \hbar$, where $N$ is the total number of particles in
the system, and $N_v$ is the number of vortices present (as large as $\sim$
300 experimentally).  The superfluid velocity, $\vec v$\,, obeys the
quantization condition,
\beq
  \oint \vec v\cdot d\vec\ell = (h/m)N_v(\cal{C}),
\eeq
where the line integral is along a closed contour surrounding
$N_v(\cal{C})$ singly quantized vortices, and $m$ is the particle mass.  The
velocity in the neighborhood of a single line is in the azimuthal direction
and has magnitude $\hbar/m\rho$, where $\rho$ is the distance from the line.
A system containing many vortex lines appears to rotate uniformly with an
average angular velocity $\Omega$, which is simply related to the
(two-dimensional) density of vortex lines, $n_v$, by the quantization
condition:  $\Omega = \pi\hbar n_v/m$.  As the system rotates more and more
rapidly, the lines become closer and closer.  What eventually happens to the
system as the angular momentum grows to the range of $10^2 N \hbar$ to
$N^2\hbar$?

    Type II superconductors in the presence of a magnetic field above a
critical value, $H_{c1}$, contain an array of vortex lines with quantized
flux.  With increasing field the line density grows until the cores begin to
overlap, at a critical field, $H_{c2}$, at which point the system turns
normal.  However, a low temperature rotating bosonic system does not have a
normal phase to which it can return.  Unlike in a weakly interacting
superconductor, where condensation occurs as a small dynamical decoration on
top of a normal state, Bose condensation occurs kinematically.  A Bose system
must respond differently than a Type II superconductor.  In a weakly
interacting atomic condensate, the radius of a single vortex core is of order
$\xi_0 = 1/\sqrt{8\pi n a_s}$, where $n$ is the particle density.  Typically,
$\xi_0 \sim 0.2$ $\mu$m, so that cores would begin to touch at
$\Omega_{c1}\sim 8\pi^2 n a\hbar/m \sim 10^3 - 10^5$ rad/sec, an
experimentally accessible rate.

    The ground state of the system rotating at angular frequency $\Omega$
about the $z$ axis is determined by minimizing the energy in the rotating
frame, $E'=E - \Omega L_z$, where $L_z$ is the component of the angular
momentum of the system along the rotation axis:
\begin{eqnarray}
 E'& = &
  \int d^3 r \left[\frac{\hbar^2}{2m} \left|\left(
   -i\nabla  -m {\vec\Omega} \times {\vec r}\right)\Psi\right|^2
   +\left(V(\vec r\,)-\frac12 m \Omega^2 r_\perp^2\right) |\Psi|^2 + \frac12 g
   |\Psi|^4\right],
\label{Eprime}
\end{eqnarray}
and $\vec r_\perp = (x,y)$.

    The fate of a rapidly rotating Bose system depends on how the system is
confined.  In typical condensate experiments the system, rotating about the
$z$ axis, is confined in a harmonic trap of the form,
\begin{equation}
  V(r_\perp,z)
= \frac12 m\left(\omega_\perp^2r_\perp^2 + \omega_z z^2\right).
\label{harmonic}
\end{equation}
In this case the centrifugal potential, $-\frac12 m \Omega^2 r_\perp^2$,
tends to cancel the transverse trapping potential, and the system cannot
rotate faster than $\omega_\perp$ without becoming untrapped.  As $\Omega \to
\omega_\perp$, the system flattens out, becomes almost two dimensional, and
eventually enters quantum Hall-like states.  The interesting physics begins in
the regime $\Omega/\omega_\perp\ga 0.9$; current experiments
\cite{jila3,coddington} have reached $\Omega/\omega_\perp\approx 0.995$.
The limit $\Omega<\omega_\perp$ is analogous to that in relativistic physics,
where particles velocities are bounded by the speed of light.  It is thus
useful to measure such rapid rotational rates in terms of a corresponding {\it
rotational rapidity} \cite{coresize}, defined by
\beq
 \tanh y = \Omega/\omega_\perp, \quad y = \frac12 \ln
\left|\frac{\omega_\perp-\Omega}{\omega_\perp+\Omega}\right|,
\eeq
which conveniently spreads out the region where $\Omega\, \la\, \omega_\perp$.

    The physics in the presence of an anharmonic transverse trapping potential
that grows faster than quadratic, e.g.,
\begin{equation}
  V_\perp(r_\perp)
  = \frac12 m\omega_\perp^2r_\perp^2(1+\lambda r_\perp^2),
 \label{anharmonic}
\end{equation}
is quite different, since, the system, being contained by the anharmonic
part of the potential, can rotate faster than $\omega_\perp$.

\begin{figure}[htbp]
\centerline{\includegraphics[height=3in]{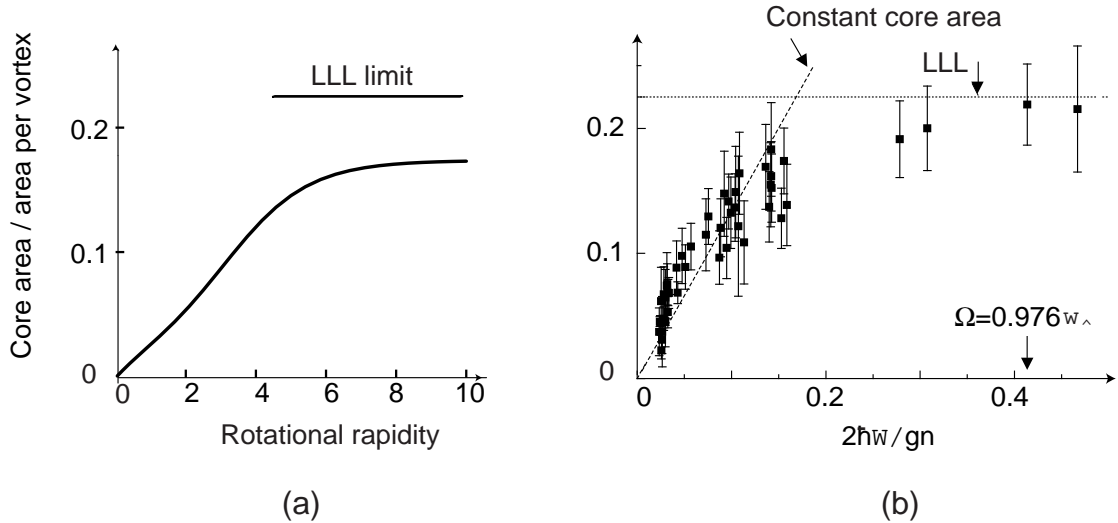}}
\caption{(a) Mean vortex core area as a fraction of the area per vortex
vs. the rotational rapidity (see text).  (b) Measured vortex core area as a
fraction of the area per vortex vs. 2$\Omega/gn$ (adapted, with permission,
from Ref.~\cite{jila3}).}
\label{coresize}
\end{figure}

\subsection{Shrinking vortex cores}

    In fact there is never a phase transition associated with the vortex cores
overlapping in a rotating Bose condensate.  Rather, as was shown in
Ref.~\cite{FG} and later \cite{coresize} in a variation treatment of the core
radius, $\xi$, the vortex cores begin to shrink as the intervortex spacing
becomes comparable to the mean field coherence length, $\xi_0$, and eventually
the core radius scales down with the intervortex spacing.  Figure 1a shows the
resultant mean square area of the vortex core, in a harmonic trap, measured in
units of the area per vortex line, for the qualitatively accurate model in
which the condensate wave function has the form $\Psi(\rho) \sim \rho$, for
$\rho<\xi$, and constant for $\xi<\rho<\ell$, where $\ell$ is the radius of
the (cylindrical) Wigner-Seitz cell around a given vortex:  $\ell^2 =
1/m\Omega$.  In this model, the mean square core area divided by the area per
vortex, $\cal{A}$, is $\xi^2/3\ell^2$.  The horizontal axis in
Fig.~\ref{coresize}a is the rotational rapidity, $y$, The linear rise at small
$\Omega$ occurs because the core size remains constant, while $\ell^2$
decreases linearly with $1/\Omega$.  The flattening of $\cal{A}$ with
increasing $\Omega$ is a consequence of the vortex radius scaling with the
intervortex spacing.  The upper line shows the exact lowest Landau level limit
(described below) as $\Omega\to\omega_\perp$.  Recent JILA measurements
\cite{jila3,coddington} of $\cal{A}$ (as a function of $2\hbar\Omega/gn$),
Fig. 1b, nicely show the expected initial linear rise, followed by the
predicted scaling of the core radius with intervortex spacing.

\subsection{Lowest Landau level regime}

    As the rotation rate in a harmonic trap approaches $\omega_\perp$, the
centrifugal potential basically cancels the transverse trapping potential; the
cloud flattens out, and becomes an effectively two dimensional system.
Because the density $n$ of the system drops, the interaction terms $\sim gn$,
become small compared with $\hbar\Omega$, Then, as one sees from
Eq.~(\ref{Eprime}) with (\ref{harmonic}), the dynamics is that of a particle
feeling the Coriolis force alone, a system formally analogous to a particle in
a magnetic field.  Ho \cite{Ho} predicted that in this limit particles should
condense into the lowest Landau level (LLL) in the Coriolis force, similar to
charged particles in the quantum Hall regime.  When $2gn \ll \Omega$, the
states in the next higher Landau level are separated by a gap $\simeq
2\omega_\perp$.  This insight has led to extensive experimental studies
\cite{HaljanCornell,jila3}.

    The single particle wave functions in the lowest Landau level have the
form,
\beq
\phi_{\mu}({\vec r_\perp}) \sim \zeta^\mu e^{-r_\perp^2/2d_{\perp}^2},
\eeq
where $\zeta = x+iy$, $\mu = 0,1,2,\dots$, and the transverse oscillator
length, $d_\perp$, is given by $\sqrt{\hbar/m\omega_\perp}$.  The LLL
condensate wave function is a linear superposition of such states:
\begin{equation}
  \Psi_{\rm LLL}({\vec r_\perp}) \sim \Sigma_\mu c_\mu \zeta^\mu
    e^{-r^2/2d_{\perp}^2} \sim \prod_{i=1}^{N}(\zeta-\zeta_i)\
   e^{-r^2/2d_{\perp}^2},
 \label{LLL}
\end{equation}
where in the latter form the polynomial $\Sigma_\mu c_\mu \zeta^\mu$ is
written as a product over its zeroes, $\zeta_i$, which are simply the
positions of the vortices in the condensate.  The state (\ref{LLL}) in the
regime $gn \ll \hbar\Omega$ is a direct continuation of the state in the
slowly rotating regime, $\hbar\Omega \ll gn$.

    As long as the total number of vortices is much larger than unity, the
energy of the cloud in the LLL limit is given by \cite{gentaro}
\begin{equation}
   E'= \Omega N
   +\int d^3r\{(\omega_{\perp}-\Omega) \frac{r_\perp^2}{d_{\perp}^2}n(\vec
    r\,) +\frac{bg}{2} n(\vec r\,)^2\},
\label{k2}
\end{equation}
plus terms involving the trapping potential in the $z$ direction.  Here
$n(\vec r\,)$ is the smoothed density profile, $=\langle |\Psi_{\rm
LLL}|^2\rangle$; the brackets denote the long wavelength smoothing.  The
energy (\ref{k2}) is minimized when the cloud assumes a density profile of the
Thomas-Fermi form \cite{rb87},
\beq
   n(r_\perp)\sim (1-r_\perp^2/R^2),
   \label{TF}
\eeq
an inverted parabola, where $R$ is the transverse radius of the cloud.
For $Na/d_z \gg 1$, where $d_z$ is the axial oscillator length, the structure
in the radial direction will be Thomas-Fermi at large $\Omega$, even if it is
Gaussian at small $\Omega$ \cite{coresize}.  In experiment
\cite{jila3,coddington}, the density profile indeed remains an inverted
parabola as $\Omega \to \omega_\perp$.

    Since the energy (\ref{k2}) depends only on the smoothed density, the
vortices must adjust their locations in order that the smoothed density be an
inverted parabola.  From the arguments in Ref.~\cite{Ho}, we find the
relation between the smoothed density and the mean vortex density,
$n_v(r_\perp)$,
\begin{equation}
 \frac{1}{4}\nabla^2 \ln n(r_\perp)  = -\frac{1}{ d_\perp^2} +\pi n_(r_\perp).
  \label{nv}
\end{equation}
For a Gaussian density profile, the vortex density is constant.  However, for
a Thomas-Fermi profile (\ref{TF}),
\begin{equation}
   n_v(r_\perp) = \frac{1}{\pi d_\perp^2} -
  \frac{1}{\pi R^2}\frac{1}{\left(1-r_\perp^2/R^2\right)^2}.
  \label{nnv}
\end{equation}
Since the second term is of order $1/N_v$ compared with the first, the
density of the vortex lattice is basically uniform (and the vortex array forms
an almost perfect triangular lattice).  Turning the argument around, very
small distortions of the vortex lattice from triangular can result in large
changes in the density distribution.  Recent measurements of the (percent
scale) distortions of the vortex lattice at relatively low rotation
rates \cite{coddington} are in good agreement with theory.

\subsection{Beyond the LLL regime}

    At sufficiently high rotation, the vortex lattice should melt and become a
vortex liquid.  The regime just beyond melting has yet to be described in
detail.  At still higher rotation speeds in harmonic traps, as seen in
numerical simulations with a limited number of particles, the system begins to
enter a sequence of highly correlated incompressible fractional quantum
Hall-like states \cite{Cooper,Viefers,Read,Jolicoeur}.  For example, at
angular momentum $L_z = N(N-1)$, where $N_v$ (measured in terms of the total
circulation, $N_v=(m/h)\oint \vec v\cdot d\vec\ell$) equals $2N$, the exact
ground state is an $N$-particle fully symmetric Laughlin wave function (in two
dimensions),
\begin{equation}
   \Psi(r_1,r_2,\dots,r_N) \sim \prod_{i\ne j}(\zeta_i-\zeta_j)^2 e^{-\Sigma_k
r_k^2/2d_\perp^2},
\end{equation}
where $\zeta_j = x_j+iy_j$.  Since the wave function vanishes whenever two
particles overlap, the interaction energy, $\frac12 g\Sigma_{i \ne
j}\delta(\vec r_i-\vec r_j)$, vanishes in this state.  Theoretically
elucidating the states in general when the angular momentum per particle is of
order the total particle number, as well as studying this regime
experimentally, remain important challenges.

\subsection{Anharmonic traps}

\begin{figure}[htbp]
\centerline{\includegraphics[height=2in]{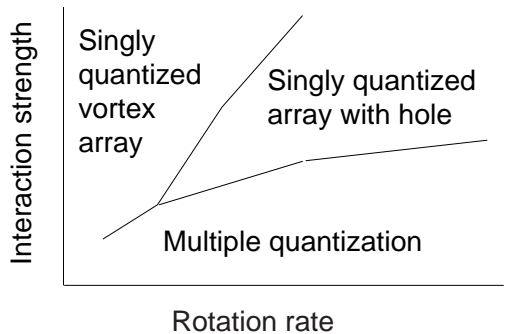}}
\caption
{Schematic phase diagram of the ground state of a rapidly rotating Bose
condensed gas at zero temperature in the $\Omega$--$N a_s/Z$ plane, where $Z$
is the height in along the rotation axis, showing the regions of multiply
quantized vortices, of singly quantized vortices forming an array which at
large number becomes a triangular lattice, and of an array of singly quantized
vortices with a hole in the center.}
\label{genphase}
\end{figure}

    The physics of a condensate confined in an anharmonic trap, e.g.,
(\ref{anharmonic}), is quite different from that in a harmonic trap, since it
becomes possible to rotate the system arbitrarily fast.  As the system rotates
sufficiently rapidly, the centrifugal force pushes the particles towards the
edge of the trap, and a hole open up in the center.  Singly quantized vortex
arrays with a hole have been seen in numerical simulations \cite{tku} and
discussed theoretically in Refs.~\cite{FG} and \cite{KB}.  In addition, at
very high rotation, systems tend to form a single multiply quantized vortex at
the center, with order parameter $\psi\sim e^{i\nu \phi}$, where the integer
quantization index $\nu$ is $>>$ 1. Such giant vortices have been seen in
numerical simulations \cite{tku,Lundh}, and are discussed theoretically in
Refs.~\cite{FG} and \cite{KB}.  The schematic phase diagram, as a function of
interparticle interaction strength vs. rotation rate is shown in
Fig.~\ref{genphase}.  Full details can be found in Refs.~\cite{KB}.  Initial
studies of rapidly rotating condensates in harmonic lattices at the ENS are
reported in Ref.~\cite{dalibard}.

\section{DEPENDENCE OF THE TRANSITION TEMPERATURE ON THE SCATTERING
LENGTH}

    The problem of determining the effects of a weak interaction, described by
an s-wave scattering length, $a_s>0$, on the Bose-Einstein condensation
transition temperature in a uniform system has had a long and tortured history
(reviewed in \cite{bigbec}) \cite{salerno}.  The problem is of great interest
to nuclear physicists since it is a system plagued by infrared divergences,
albeit not as severe as those encountered in quark-gluon plasmas.  The result
for the transition temperature, as finally resolved in \cite{bigbec}, is that
for small $a_sn^{1/3}$ the shift in the transition temperature is linear in
$a_s$ and positive:  \beq \frac{\Delta T_c}{T_c^0} = c\, a_sn^{1/3},
\label{deltat} \eeq where $T_c^0$ is the free gas transition temperature,
given by $\zeta(3/2)\lambda^{-3}$, with thermal wavelength $\lambda = (2\pi
\hbar^2/m\kappa_B T)^{-1/2}$, and $T_c = T_c^0 + \Delta T_c$ is the transition
temperature in the presence of interactions.  The positive constant $c$ is of
order 1-2, but cannot be calculated in closed form.  Because the low energy
s-wave interparticle interaction, $\sim 4\pi \hbar^2 a_s/m$, is first order in
$a_s$, it would appear at first glance that the result (\ref{deltat}) should
be straightforward to derive.  However, the first order, or mean field shift,
in the particle energies has no effect on the transition temperature, since it
is independent of momentum and thus compensated by a shift in the chemical
potential.  In second and higher order in the dimensionless expansion
parameter $a_s/\lambda$, each and every term in the perturbation expansion
involves integrals that diverge in the infrared, implying that to find a
finite answer one has to sum an infinite set of terms.  The problem is a true
critical phenomenon, completely outside the range of Gross-Pitaevskii mean
field.

    Successful approaches to the problem have been three-fold.  The first is
via numerical simulation of finite systems, including path integral
Monte-Carlo calculations \cite{gruter,holz}, and more recent calculations of
$\phi^4$ field theory on the lattice, extrapolated to the continuum limit
\cite{arnold,kash}.  The second has been through analysis of the scaling
structure of the perturbation expansion in finite temperature field theory
\cite{bigbec}.  A third approach has been by means of the renormalization
group, carried out for a field theory with $N$ components (compared to the two
components corresponding to the real and imaginary parts of the order
parameter $\Psi$) in \cite{bbz}; in the limit where the number of components
goes to $\infty$ the exact result for the shift in the transition temperature
is $c= 8\pi/3\zeta(3/2)^{4/3} = 2.33$.  In a trapped Bose gas, the finite
level spacing regulates the infrared divergences; $T_c$ decreases because the
interparticle interactions lower the density.  With interaction effects, as
calculated by Arnold and Tom\'a{\^s}ik \cite{arnoldtrap}, the relative shift
is $\sim 3.4 a_s/\lambda$, where $\lambda$ measures the interparticle spacing
in the center of the trap.

    Interestingly, the various calculations of $c$ in a homogeneous gas differ
noticeably, ranging from $0.34\pm 0.06$ \cite{gruter} to $\simeq$ 1.3
\cite{arnold,kash}), to $2.33 \pm 0.25$ \cite{holz}.  The discrepancy has a
physical origin, namely that the expansion expansion of $\Delta T_c/T_c^0$ in
powers of $a_s/\lambda$ has a highly non-analytic structure.  In fact, the
series cannot be analytic, since an infinite homogeneous system of bosons with
negative $a_s$ is unstable against collapse.  Were the series analytic about
zero scattering length, it would have to hold for sufficiently small $|a_s|$,
whether $a_s$ is positive or negative.  Rather, the series is asymptotic for
positive small $a_s$.  As shown in \cite{nlo}, to next leading order the
series has the form
\beq
  \frac{\Delta T_c}{T_c^0}= c\, (a_sn^{1/3}) + d\, (a_sn^{1/3})^2 \ln
(a_sn^{1/3}) + {\cal O}\left(((a_sn^{1/3})^2\right),
\eeq
with $d$ positive.  The logarithmic term introduces rapid variation of
$\Delta T_c/T_c^0$ with $a_s$.  In the limit of large $N$,
\beq
\frac{\Delta T_c}{T_c}= \frac{8\pi}{3\zeta(3/2)}\frac{a_s}{\lambda}
 \left\{
  1 + 16 N \frac{a_s}{\lambda^2 \Lambda}
   \ln \frac{N as}{\lambda^2 \Lambda}
     + {\cal O}\left( \frac{N as}{\lambda^2 \Lambda} \right)
   \right\},
\label{tcres}
\eeq
where $\Lambda \sim 1/\lambda$ is the effective ultraviolet cutoff in a
calculation within classical field theory near $T_c$.  The coefficient of the
logarithmic term has been calculated exactly by Arnold et al.
\cite{arnoldlog}, who find $d\simeq$ 19.7518; the logarithmic term in
Eq.~(\ref{tcres}) for $N=2$ and $\Lambda \sim \sqrt{2\pi}/\lambda$ is in
reasonable agreement with this exact result.  The logarithmic term reduces the
shift by about a factor of two for $a_sn^{1/3} \sim 10^{-2}$ and by a factor
of order six for $a_sn^{1/3} \sim 10^{-2}$, thereby substantially improving
the agreement with the calculations of Ref.  \cite{gruter}, for which the
lowest data point was at $a_sn^{1/3} \sim 10^{-2}$.  Quite generally, the
logarithmic corrections play an important role in comparison of the various
numerical calculations.

\section{FESHBACH RESONANCES AND UNIVERSALITY}

    One enormous advantage enjoyed by experimentalists working with trapped
atomic systems is the ability to vary the parameters of the system over ranges
inconceivable in other systems.  One can readily control the type of atoms,
their number and density, and the shape of the trap.  Optical lattices permit
one to study systems in varying dimensions and confinement.  Perhaps the most
spectacular freedom of these systems is the ability to control the strength
and sign of the interactions between the particles, via Feshbach resonances.

\begin{figure}[htbp]
\centerline{\includegraphics[height=1.2in]{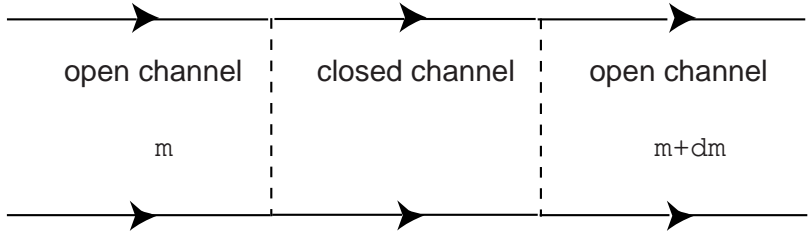}}
\caption
    {Two atoms in the initial (open) channel, with total magnetic moment
$\mu$, scattering through an intermediate state (closed channel) with total
magnetic moment $\mu +\delta\mu$.  When the energies of the intermediate and
initial states are tuned via a magnetic field, the scattering has a Feshbach
resonance.
}
\label{feshbach}
\end{figure}

\begin{figure}[htbp]
\centerline{\includegraphics[height=3.5in]{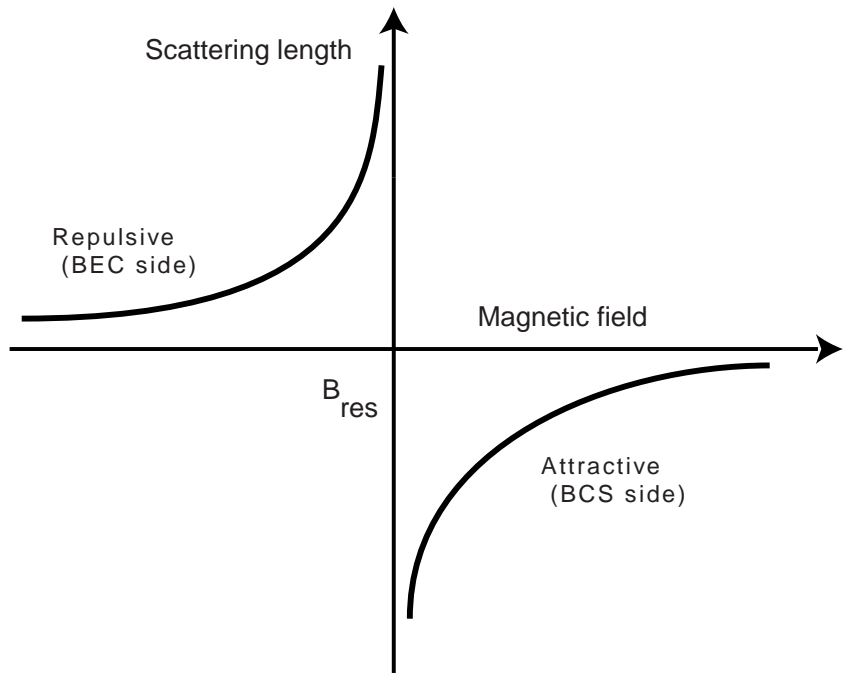}}
\caption
    {Scattering length vs. magnetic field in neighborhood of a Feshbach
resonance at magnetic field $B_{res}$.
}
\label{feshbach1}
\end{figure}

    In the scattering of two atoms, total angular momentum is conserved.
However, the atoms, through the hyperfine interaction, can scatter into
intermediate states (Fig.~\ref{feshbach}) in which one or both atoms are in
different hyperfine states.  Since the magnetic moment of the atomic states is
determined by the valence electron configuration, the net magnetic moment,
$\mu + \delta\mu$, in the intermediate states can be different from that of
the initial state, $\mu$.  Thus in the presence of a magnetic field, $B$, the
relative energies of the initial (0) and intermediate (i) states will be
shifted by $\delta\mu B$.  If the intermediate and initial levels cross at a
magnetic field, $B_{res}$, the two particle scattering amplitude will have a
resonance, known as a Feshbach resonance, there.  The process in
Fig.~\ref{feshbach} gives a contribution to the scattering amplitude
\beq
     \delta a_s \sim  \frac{|M|^2}{E_0(B)-E_i(B)},
\eeq
so that $a_s$ near resonance has the form,
\beq
      a_s(B) = a_b\left(1 - \frac{\Delta}{B-B_{res}}\right),
\eeq
as shown schematically in Fig.~\ref{feshbach1}, where the background
scattering length $a_b$ is the value of the scattering length away from
resonance.

    Note that not only does the scattering amplitude diverge at the resonance,
it changes sign across the resonance.  Thus for the case illustrated in
Fig.~\ref{feshbach1} (for $\Delta>0$, as occurs in both stable fermion
alkalis, $^6$Li and $^{40}$K) the effective interaction between the atoms
changes from repulsive for $B< B_{res}$ to attractive for $B > B_{res}$.  To
understand the basic physics a useful first picture is to think in terms of
the particles interacting via an {\it attractive} short ranged potential with
an s-wave bound state near zero energy.  If the bound state is just below
zero, then the s-wave scattering length is positive, so that particles at low
energy feel a residual repulsion.  This situation corresponds to the region
in Fig.~\ref{feshbach1} to the left of the resonance.  Decreasing the depth
of the potential is equivalent to moving to the right in this figure.  As the
potential becomes less deep and the bound state approaches zero energy, the
scattering length grows, and finally as the bound state moves into the
continuum, the scattering length becomes negative, and particles in the
continuum attract.

    Early experiments employed the Feshbach resonance in bosonic $^{85}$Rb, at
B=155G, to study, e.g., how a cloud initially with a positive scattering
length undergoes collapse as the interaction is suddenly made attractive
\cite{cornish}.  Currently, Feshbach resonances are being employed to bring
Fermi systems into the strongly interacting regime, where one can induce BCS
pairing, as well as study the crossover from a Bose condensate of two-fermion
molecules to BCS superfluidity.

\subsection{Physics at resonance}

    Away from a Feshbach resonance, the interparticle spacing is generally
large compared with the scattering length, i.e., $na_s^3\ll 1$, where $n$ is
the density, so that the systems can be treated as weakly interacting.
Furthermore, the range of the actual interatomic potentials is small compared
with either $a_s$ or $n^{-1/3}$, so that one has a clean separation of length
scales.  In the neighborhood of a Feshbach resonance, however, the scattering
length becomes large compared with the interparticle spacing, $a_s\gg
n^{-1/3}$, and the system enters a strongly interacting {\it unitarity}
regime.  Here the scattering length is an irrelevant parameter, as is the
microscopic scale of the interatomic potential.  The two-particle s-wave cross
section is unitarity limited by 8$\pi/k^2$, where $k$ is the relative momentum
of the colliding pair.  Thus the only relevant length scale in the problem is
the interparticle spacing, and the structure of the system should exhibit
universality.

    In particular, in a system of bosons or fermions, the ground state energy
per particle must have the form, $E/N \sim n^{2/3}/m$; one can write,
\beq
 E/N = \frac35 E_f^0 (1+\beta),
\eeq
where in a Fermi gas the free particle Fermi energy, $E_f^0$ equals
$k_f^2/2m$, with $k_f$ the Fermi momentum, and $\beta$ is a universal
parameter (dependent only on statistics) and non-trivial to calculate.  Via a
fixed node Monte Carlo method, Carlson et al.~\cite{carlson} derive $\beta =
-0.56$, while using a lowest order constrained variational calculation,
Heiselberg \cite{hh} finds $\beta = -0.33$.  These results, and notably the
sign of $\beta$, are in reasonable agreement with experimental determinations
of the energy via studying the expansion of a cloud released from the trap,
e.g., \cite{thomas-energy,ens-energy}.

    The expansion of a cloud in the resonant regime exhibits elliptic flow
\cite{thomas-flow}, quite similar to that observed in ultrarelativistic heavy
ion collisions.  Indeed strongly coupled quark-gluon plasmas and strongly
coupled atomic fermionic clouds share the feature that both appear to be very
nearly perfect fluids \cite{edward}.

\section{FERMION PAIRING}

    S-wave pairing in a fermion system, e.g., electrons, or nucleons, normally
takes place between particles of opposite spin.  However, magnetic traps
permit one to trap only the low-field seeking states.  Furthermore, the
temperatures one can reach are generally too large a fraction of the Fermi
temperature for BCS pairing to take place without enhancing the interaction
strength.  Thus the strategy to achieve pairing in a magnetic trap is to
populate equally two hyperfine atomic states, and by means of a Feshbach
resonance, strengthen the interaction between them \cite{res-sc}.  (Note that
owing to the Pauli principle, interactions within a given hyperfine level are
very small, since the interactions are very short ranged.)

    In the regime to the right of the resonance in Fig~\ref{feshbach1}, where
the interactions between particles in the continuum are attractive, one
expects fermions in the two hyperfine levels to pair, with a transition
temperature
\beq
 T_c \sim T_f e^{-1/k_f|a_s|},
\eeq
with $T_f$ is the Fermi temperature.  As one moves towards the resonance,
$T_c$ grows.  To the left of the resonance, pairs of atoms can fall into a
weakly bound state, forming molecules; in this regime, one expects the system
to consist of a Bose-Einstein condensate of these molecules.  In fact, as one
goes through the resonance starting from the right, the Cooper pairs in the
BCS state decrease in size, and go continuously into the Bose-Einstein
condensate of molecules.  Bose-Einstein condensation of di-fermion molecules
and BCS pairing are two ends of a continuum with no phase transition en route
\cite{eagles,tony,nsr}.  The trapped atomic systems give one the first
opportunity to study this BEC-BCS crossover experimentally.  The situation is
remarkably similar to that discussed in color superconductivity of a
quark-gluon plasma, where the color-flavor-locked state goes continuously,
with decreasing baryon density, into a gas of nucleons \cite{colorsup}.  The
precise analogue would be the crossover in color-SU(2) from paired quarks at
high density to Bose-condensed mesons at lower density \cite{su2}.

    Experiments to produce paired fermions proceed by cooling a gas of
fermions in the high magnetic field state, where the interactions are weakly
attractive, but the temperature is too high to have BCS pairing.  By ramping
the magnetic field adiabatically to below the resonance one can slide the
fermions into the weakly bound state there, producing a gas of long-lived
Bose-condensed molecules \cite{randy,regal,ens-mol,jochim,zwier2}.  Such
coherent molecule production is reversible, with no entropy generation.  The
system passes in fact through the strongly coupled BCS regime to the right of
the resonance.  To detect this state, Regal et al.  \cite{jin} at JILA
(starting with $^{40}$K gas at initial temperature, $T/T_f \sim 0.08$) stopped
the adiabatic ramp just to the right of the resonance; they then carried out a
rapid ramp across the resonance, which effectively projects the fermions onto
molecules.  By measuring the momentum distribution of these molecules they
then infer the existence of a fermion condensate in the state to the right
of the resonance.  Similar production of paired fermions has been carried out
at Duke \cite{kinast}, MIT \cite{zwierlein}, and Innsbruck \cite{chin}.
Theoretical predictions of the phase diagram near the BEC-BCS transition
\cite{diener} agree with measurements.

\subsection{Direct measurements of pairing}

    A very important issue is how to detect pairing and superfluidity
directly in ultracold Fermi gases.  It should be stressed that these are
distinct questions, since one can have pairing of fermions into molecules that
are not condensed.  Direct measurements of superfluid behavior include
production of stable vortices, or other persistent current phenomena, the
Josephson effect, and reduction of moments of inertia, as in superfluid helium
II and nuclei with nucleon pairing.  Such experiments have yet to be carried
out.

\begin{figure}[htbp]
\begin{center}\vspace{0.0cm}
\rotatebox{0}{\hspace{-0.cm}
\resizebox{10cm}{!}
{\includegraphics{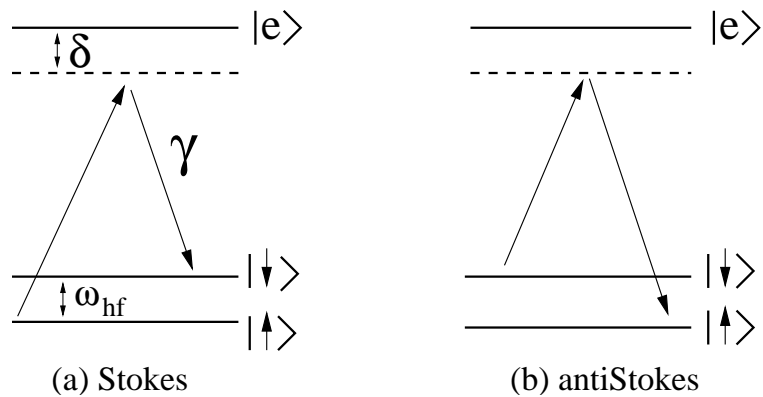}}}
    \caption{(a) Stokes and (b) anti-Stokes scattering of light between two
hyperfine states, as a signal of BCS pairing between the states.}
\label{Stokes}
\end{center}
\end{figure}

\begin{figure}[htbp]
\begin{center}\vspace{0.0cm}
\rotatebox{0}{\hspace{-0.cm}
\resizebox{9cm}{!}
{\includegraphics{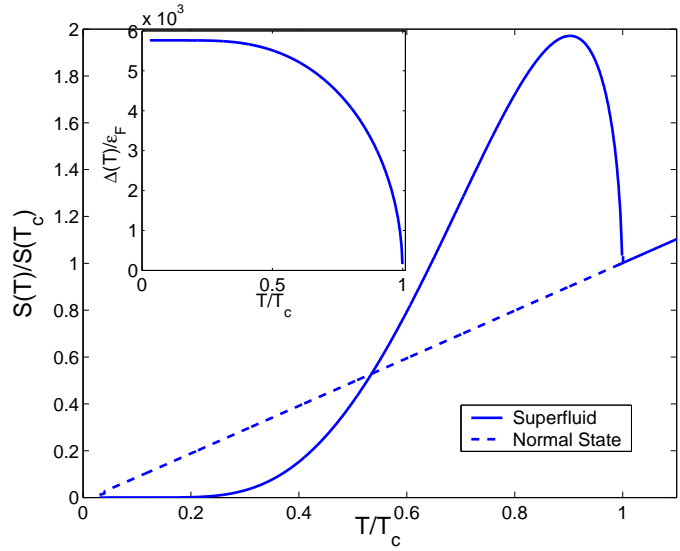}}}
\caption{Scattered light intensity as a function of temperature showing
the Hebel-Slichter enhancement below $T_c$.  The inset shows the BCS gap
$\Delta(T)$.}
\label{Slichter}
\end{center}
\end{figure}

\begin{figure}[htbp]
\begin{center}\vspace{0.0cm}
\rotatebox{0}{\hspace{-0.cm}
\resizebox{9cm}{!}
{\includegraphics{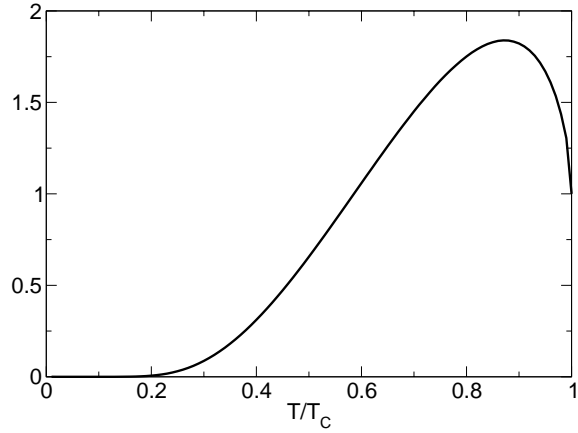}}}
\caption{Cross section for scattering of neutrinos from superfluid nuclear
matter, in units of the normal state cross section, showing the Hebel-Slichter
enhancement below $T_c$.  I am very grateful to Sanjay Reddy for this figure.
}
\label{reddy}
\end{center}
\end{figure}

    Pairing can detected by measurement, in principle, of the energy gap,
$\Delta$, entering the quasiparticle energy,
\beq
  E = \sqrt{(\epsilon - \mu)^2 + \Delta^2},
\eeq
where $\epsilon$ is the energy in the absence of pairing and $\mu$ is here
the chemical potential.  For example, Chin et al.  \cite{chin}, have trapped
$^6$Li above a Feshbach resonance at magnetic field $\sim 830$G, with equal
populations of atoms in the nuclear spin states $m_I = -1$ and 0. At
sufficiently low temperatures the atoms in the two states should be s-wave
paired.  By rf they excite an atom from $m_I=0$ to 2, and find a peak in the
spectrum at higher energy than would be seen for unpaired atoms; the extra
energy is interpreted as that required, $\ge 2\Delta$, for breaking a Cooper
pair.  A second method of detecting pairing is through measurement of the
specific heat, which shows a depression at low temperatures due to the finite
gap in single particle excitation spectrum \cite{thomas-cv}.  The theoretical
interpretation of these measurements is given in \cite{levin}.  Changes in
collective mode spectra have also been interpreted as evidence for the onset
of superfluidity \cite{kinast,bart}.

    Arguably the most crucial test originally of the BCS theory of
superconductivity was the Hebel-Slichter effect which demonstrated a large
enhancement of the nuclear spin relaxation rate, due to coupling to paired
electrons, at temperatures just below the superconducting transition
temperature, $T_c$ \cite{Hebel}.  The peak occurs in the spin-flip spin-flip
correlation function (the nuclear axial current--axial current correlation
function).  The analogous effect in pairing of fermions between two hyperfine
levels (denoted here by $|\uparrow\rangle$ and $|\downarrow\rangle$) can be
seen via laser excitation from $|\uparrow\rangle$ to a high-lying level,
$|e\rangle$, off-resonance, which then de-excites to $|\downarrow\rangle$,
with emission of a photon.  The two possible processes are illustrated in
Fig.~\ref{Stokes}.  In the first (a) the energy of the emitted photon is less
than that of the incident (Stokes scattering) and in the second (b) the
emitted photon has higher energy (anti-Stokes).  Figure~\ref{Slichter} shows
the expected enhancement in the rates at temperature for a weakly coupled
paired system just below the superfluid transition; the enhancement is a very
clear signal of the onset of pairing.  Interestingly, the same effect occurs
in the scattering of neutrinos from superfluid nuclear and color
superconducting matter, Fig.~\ref{reddy}, which involves the analogous nuclear
axial current correlation function \cite{sanjay}.

      This research was supported in part by US National Science Foundation
Grants PHY00-98353 and PHY03-55014.

\end{document}